\documentclass[10pt, onecolumn]{IEEEtran}

\usepackage{balance} 

\usepackage{xspace}
\usepackage[normalem]{ulem}

\usepackage{enumitem}
\usepackage{tabularx}
\usepackage{multirow}
\usepackage{color}
\usepackage{booktabs}

\usepackage{cite}
\usepackage{amsmath,amssymb,amsfonts}
\usepackage{mathtools}
\usepackage[pdftex]{graphicx}
\usepackage{textcomp}
\graphicspath{{./Figures/}}
\DeclareGraphicsExtensions{.pdf,.jpeg,.png}

\usepackage{algorithm}
\usepackage{algpseudocode}

\newif\ifcomment
\commentfalse

\newcommand{\RW}[1]{{\noindent \color{red}#1}}

\newcommand{\cps}{CPS\xspace}

\newcommand{\ms}{\ensuremath{\,\text{ms}}\xspace}
\newcommand{\us}{\ensuremath{\,\mu\text{s}}\xspace}

\newcommand{\taskset}{\ensuremath{\mathcal{T}}\xspace}
\newcommand{\messageset}{\ensuremath{\mathcal{M}}\xspace}

\newcommand{\predG}{\ensuremath{\mathcal{G}}\xspace}
\newcommand{\beacon}{\ensuremath{b}\xspace}

\newcommand{\sched}[1]{\ensuremath{Sched(#1)}\xspace}

\newcommand{\appl}[1]{\ensuremath{\mathit{a_{#1}}}\xspace}
\newcommand{\appi}{\appl{i}}
\newcommand{\appj}{\appl{j}}
\newcommand{\app}{\appl{}}

\newcommand{\mode}[1]{\ensuremath{\mathit{M_{#1}}}\xspace}
\newcommand{\modei}{\mode{i}}
\newcommand{\modej}{\mode{j}}

\newcommand{\Tround}{\ensuremath{T_{r}}\xspace}
\newcommand{\Troundon}{\ensuremath{T_{r}^{\mathit{on}}}\xspace}
\newcommand{\Tworound}{\ensuremath{T_{\mathit{wo/r}}}\xspace}
\newcommand{\Tworoundon}{\ensuremath{T_{\mathit{wo/r}}^{\mathit{on}}}\xspace}

\newcommand{\Tslot}{\ensuremath{T_{\mathit{slot}}}\xspace}
\newcommand{\Twakeup}{\ensuremath{T_{\mathit{wake-up}}}\xspace}
\newcommand{\Tstart}{\ensuremath{T_{\mathit{start}}}\xspace}
\newcommand{\Tglossy}{\ensuremath{T_{\mathit{flood}}}\xspace}
\newcommand{\Thop}{\ensuremath{T_{\mathit{hop}}}\xspace}
\newcommand{\Td}{\ensuremath{T_{d}}\xspace}
\newcommand{\Tcal}{\ensuremath{T_{\mathit{cal}}}\xspace}
\newcommand{\Lcal}{\ensuremath{L_{\mathit{cal}}}\xspace}
\newcommand{\Theader}{\ensuremath{T_{\mathit{header}}}\xspace}
\newcommand{\Lheader}{\ensuremath{L_{\mathit{header}}}\xspace}
\newcommand{\Tpayload}{\ensuremath{T_{\mathit{payload}}}\xspace}
\newcommand{\Rbit}{\ensuremath{R_{\mathit{bit}}}\xspace}
\newcommand{\Tgap}{\ensuremath{T_{\mathit{gap}}}\xspace}
\newcommand{\Ton}{\ensuremath{T^{\mathit{on}}}\xspace}
\newcommand{\Toff}{\ensuremath{T^{\mathit{off}}}\xspace}
\newcommand{\Lbeacon}{\ensuremath{L_{\mathit{beacon}}}\xspace}

\renewcommand{\prec}{\ensuremath{\mathit{prec}}\xspace}
\newcommand{\map}{\ensuremath{\mathit{map}}\xspace}

\newcommand{\af}{\ensuremath{\mathit{af}}\xspace}
\newcommand{\df}{\ensuremath{\mathit{df}}\xspace}
\renewcommand{\sf}{\ensuremath{\mathit{sf}}\xspace}
\newcommand{\id}{\ensuremath{\mathit{id}}\xspace}
\newcommand{\ids}{\ensuremath{\mathit{id}}s\xspace}
\newcommand{\SB}{\ensuremath{\mathit{SB}}\xspace}
\newcommand{\first}{\ensuremath{\mathit{first}}\xspace}
\newcommand{\last}{\ensuremath{\mathit{last}}\xspace}
\newcommand{\obj}{\ensuremath{\mathit{obj}}\xspace}

\newcommand{\name}{\textsc{TTW}\xspace}

\newcommand{\fakepar}[1]{\vspace{1mm}\noindent\textbf{#1.}}

\newcommand{\eg}{\emph{e.g.},\xspace}
\newcommand{\ie}{\emph{i.e.},\xspace}

\newcommand{\capt}[1]{\mdseries{\emph{#1}}}

\newcommand\figref[1]{Fig.\,\ref{#1}}
\newcommand\secref[1]{Sec.\,\ref{#1}} 

\begin{document}

\title{\name: A Time-Triggered-Wireless Design for \cps\\
{\Large [ Extended version ]}}

\author{\IEEEauthorblockN{%
Romain Jacob\IEEEauthorrefmark{1}\hfill
Licong Zhang\IEEEauthorrefmark{2}\hfill
Marco Zimmerling\IEEEauthorrefmark{3}\hfill
Jan Beutel\IEEEauthorrefmark{1}\hfill
Samarjit Chakraborty\IEEEauthorrefmark{2}\hfill
Lothar Thiele\IEEEauthorrefmark{1}}\\
\IEEEauthorblockA{%
\begin{minipage}{0.3\linewidth}%
\centering
\IEEEauthorrefmark{1}ETH Zurich, Switzerland\\
firstname.lastname@tik.ee.ethz.ch
\end{minipage}
\hfill
\begin{minipage}{0.3\linewidth}%
\centering
\IEEEauthorrefmark{2}TU Munchen, Germany\\
firstname.lastname@rcs.ei.tum.de
\end{minipage}
\hfill
\begin{minipage}{0.3\linewidth}%
\centering
\IEEEauthorrefmark{3}TU Dresden, Germany\\
marco.zimmerling@tu-dresden.de
\end{minipage}}}

\maketitle


\begin{abstract}
Wired field buses have proved their effectiveness to support Cyber-Physical Systems (\cps).
However, in avionics, for ease of deployment, or for new functionality featuring mobile devices, there is a strong interest for wireless solutions.
Low-power wireless protocols have been proposed, but requirements of a large class of CPS applications can still not be satisfied.
This paper presents Time-Triggered-Wireless (\name), a distributed low-power wireless system design that minimizes energy consumption and offers end-to-end timing predictability, adaptability, reliability, low latency.
Our evaluation shows a reduction of communication latency by a factor 2x and of energy consumption by 33-40\% compared to state-of-the-art approaches. 
This validates the suitability of \name for wireless \cps applications and opens the way for implementation and real-world experience with industry partners.

\end{abstract}



\section{Introduction}\label{sec:intro}


Commonly, \cps are understood as systems where \emph{``physical and software components are deeply intertwined, each operating on different spatial and temporal scales, exhibiting multiple and distinct behavioral modalities, and interacting with each other in a myriad of ways that change with context''}~\cite{NSF10}.
Application domains include robotics, distributed monitoring, process control, and power-grid management.
\cps combine physical processes, sensing, online computation, communication, and actuation in a single distributed control system.

To support \cps, industry has been widely relying on wired field buses -- CAN, FlexRay, ARINC 429, AFDX in automotive and avionics -- with good reasons. They combine functional and non-functional predictability with appropriate bandwidth, message delay, and fault-tolerance.
Yet, several of the above application domains would benefit from wireless communication for its ease of installation, logical and spatial reconfigurability, and flexibility. In other cases, wireless is the only option, due to the presence of mobile devices for example.

\fakepar{Challenges}
One major obstacle in wireless communication has been the reliability of packet transmission.
Recently, several low-power protocols featuring very low packet loss rate have been proposed and partly standardized, based on the principles of time-division-multiplex and time-slotted access, see TSCH~\cite{TSCH}, WirelessHART~\cite{hart}, and LWB~\cite{ferrari2012low}. Dependability competitions where such protocols have been tested in high-interference environments~\cite{2017_EWSNCompetition} provide additional evidence.

Despite these achievements, the available protocols do not yet satisfy all requirements of typical CPS applications, \ie reliability, timing predictability, low end-to-end latency at the application level, energy efficiency, and quick runtime adaptability to different modes of operation~\cite{akerberg11}.

To understand the challenges of wireless CPS, it is helpful to realize the fundamental difference between a field bus and a wireless network. In a field bus, whenever a node is not transmitting, it can idly listen for incoming messages. Upon request from a central host, each node can \emph{wake-up and react quickly}. For a low-power wireless node, the major part of the energy is consumed in its radio. Therefore, energy efficiency requires to turn the radio off whenever possible to enable long autonomous operation without an external power source. A node is then \emph{unreachable until it wakes up}. Thus, two nodes require overlapping wake-up time intervals to communicate.

This observation often results in wireless system designs that minimize energy consumption by using \emph{rounds}, \ie time intervals where all nodes wake-up, exchange messages, then turn off their radio, see~\cite{hart,ferrari2012low,TSCH}. Scheduling policies define when the rounds take place -- \ie when to wake up -- and which nodes are allowed to send messages during the round.
However, \cps also execute tasks, \eg sensing or actuation. Oftentimes, the system requirements are specified end-to-end, \ie between distributed tasks exchanging messages. Meeting such requirements calls for co-scheduling the execution of tasks and the transmission of messages, as proposed in the literature for wired architectures, see~\cite{abdelzaher1999combined,craciunas2016coscheduling,ashjaei2017end2endReservation}.

Following the above discussion, we focus on round-based wireless designs. Three challenges arise.
First, state-of-the-art scheduling methods for wired buses \emph{are not directly applicable}. They assume that communication can be scheduled at any point in time, which does not conform to the concept of rounds in a wireless setting.
Second, the optimization problem for co-scheduling tasks and messages \emph{cannot be solved online} in a low-power setting; pre-computed schedules are highly desirable in hard real-time systems.
Finally, \cps often \emph{require runtime adaptability} while the systems must remain safe when considering packet loss and mode changes.

\fakepar{Contributions}
To address these challenges, we propose models and methods enabling low-power wireless communication suited to \cps, based on communication rounds, and aiming for end-to-end guarantees and adaptability properties comparable to a wired bus.
We partition the co-scheduling problem into offline and online phases.
All schedules are synthesized offline and distributed to the network at deployment time.
At runtime, the current mode of operation and the schedule phase are broadcast to all participating nodes (see \secref{sec:implementation} for details).

This approach induces very little protocol communication overhead, which is of importance in low-power settings.
Moreover, a beacon at the beginning of each round enables reliable execution of the protocol, even under packet loss and mode changes.
In summary, we make the following contributions.
\begin{enumerate}[leftmargin=*, topsep=0pt]
\item We present Time-Triggered-Wireless (\name), a novel low-power wireless system design that meets common requirements of distributed CPS applications such as reliability, timing predictability, low end-to-end latency at the application level, energy efficiency, and quick runtime adaptability to different modes of operation.
	
\item We formally specify the joint optimization problem of co-scheduling distributed tasks, messages, and communication rounds which guarantees timing, minimizes end-to-end latency between application tasks, minimizes energy, and ensures safety in terms of conflict-free communication even under packet loss. We provide a methodology that efficiently solves this optimization problem, known to be NP-hard~\cite{jeffay1991NPnonpreemptive}.
	
\item Using time and energy models, we validate the benefits of rounds in a wireless design to minimize energy, and we derive the minimum end-to-end latency achievable.	
	
\end{enumerate} 

\section{\name System Design}\label{sec:implementation}

Before delving into technical details, this section presents the underlying concepts of Time-Triggered-Wireless (\name) with some details on the overall system design.
Then \secref{sec:model} and \secref{sec:ILP} will introduce the formal application model and our solution to the corresponding scheduling problem.

\subsection{\name Foundations}

\begin{figure}
\begin{minipage}[b]{0.27\linewidth}
\centering
\includegraphics[height=90pt]{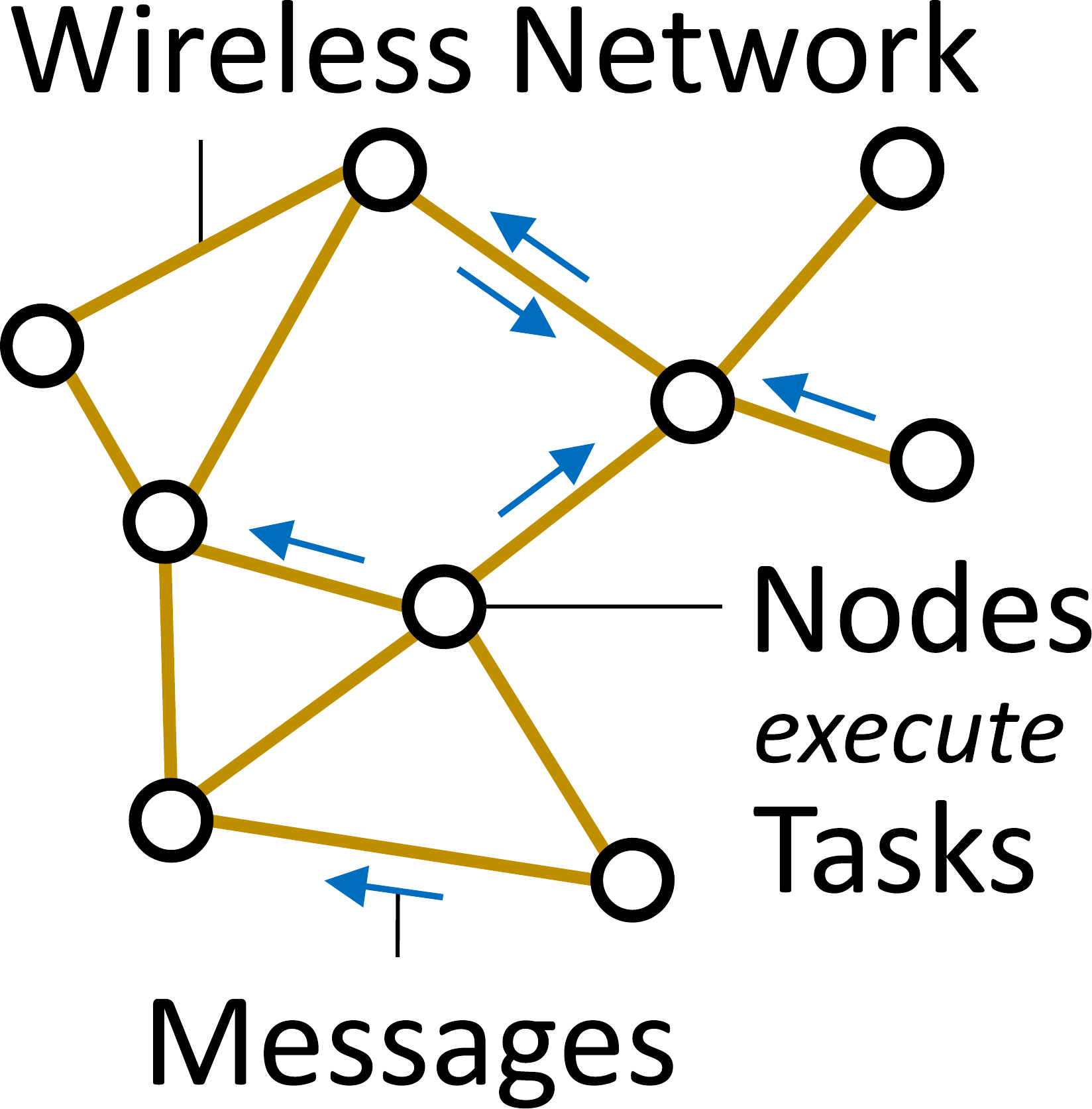}\\
{\footnotesize (a)}
\end{minipage}
\hfill
\begin{minipage}[b]{0.7\linewidth}
\centering
\includegraphics[height=90pt]{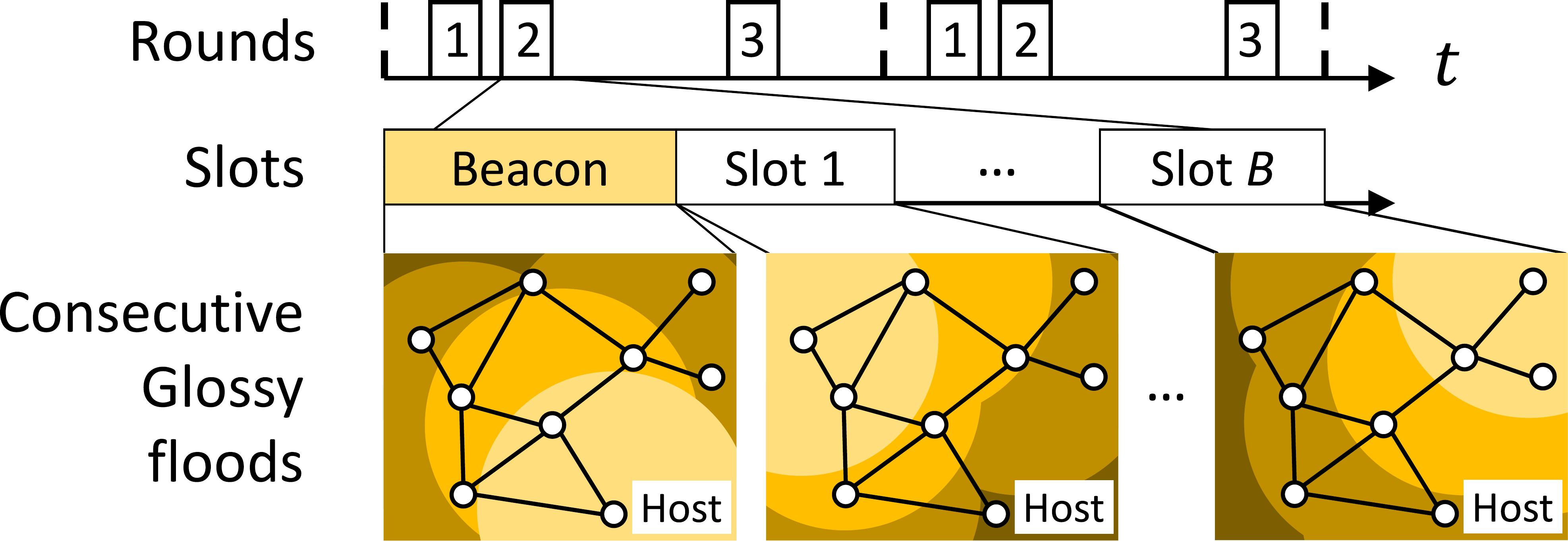}\\
{\footnotesize (b)}
\end{minipage}%
\vspace{-0.15cm}
\caption{(a) General system model and (b) time-slotted execution of \name.
\capt{%
As in LWB\emph{~\cite{ferrari2012low}}, communication rounds are divided into time slots, in which Glossy floods are executed. Each color shows one flooding step.
In \name, the first slot of each round contains a beacon sent by the host, followed by (up to) $B$ slots, allocated to application messages. The beacons announce the identification number of the round (round \id) and trigger mode changes.
}}
\vspace{-0.3cm}
\label{fig:overview}
\end{figure}

We consider a set of \emph{nodes} connected by a \emph{wireless multi-hop network}, as illustrated in \figref{fig:overview}(a).
Distributed \emph{applications}, composed of multiple \emph{tasks}, are executed on the network. Tasks are mapped to nodes and exchange \emph{messages} wirelessly.
To minimize the energy consumed for wireless communication, we \emph{group message transmissions into communication rounds}, \ie time intervals where all nodes turn their radio on and communicate. 
Rounds are composed of (up to) $B$ \emph{contention-free slots} -- each slot is allocated to a node, which is the only one allowed to start transmitting during this time slot.
%
%
The network is controlled centrally by a node called the \emph{host}. 
Commands are sent by the host at the beginning of each round in an additional slot, called \emph{beacon}.
Within each slot, communication is realized by \emph{network-wide Glossy floods}~\cite{ferrari11}. 
Rounds, slots and floods are illustrated in \figref{fig:overview}(b).

Those concepts were taken from the Low-power Wireless Bus (LWB)~\cite{ferrari2012low}, another protocol that inspired the design of \name. LWB has a number of benefits.
It is based on Glossy, which has been proven to be highly reliable, energy efficient (see~\cite{2017_EWSNCompetition})
and provides sub-microsecond time synchronization accuracy~\cite{ferrari11}.
As flooding in Glossy is independent of the network state, it creates a \emph{virtual single-hop network}, where every node can directly communicate with every other node.
This enables to schedule LWB \emph{as a shared bus}.
Finally, as each message is received by all nodes, LWB seamlessly supports unicast, multicast and broadcast transmissions.
For a given payload, the transmission time only depends on the network diameter, \ie the maximal hop distance between two nodes. 

However despite those benefits, LWB is unable to solve the wireless \cps challenge. 
	It does not account for the scheduling of distributed tasks
	and 
	it does not provide any timing guarantees.
This motivates the design of \name, a novel wireless system design which co-schedules tasks and messages.

\subsection{Building-up \name}

Building on those foundations, we present the concepts that create \name, a solution satisfying the requirements of \cps applications formulated in the introduction.

\fakepar{Low end-to-end latency}
To achieve low end-to-end latency%
	\footnote{\eg 10-500 \ms delay for a distributed closed-loop control system~\cite{akerberg11}}%
, \name \emph{co-schedules task executions and message transmissions}%
\footnote{similar to the state of the art for wired protocols~\cite{craciunas2016coscheduling,zhang14}}%
\emph{, together with the communication rounds.}

Such scheduling problem is a complex optimization that cannot be solved on-line, even less in a low-power setting. 
Therefore, \name \emph{statically synthesize the schedule} of all tasks, messages, and rounds to meet real-time constraints, minimize end-to-end latency, and minimize the energy consumed for communication. However, because of the dependencies between messages and rounds, conventional methods cannot be applied directly. A novel modeling approach is required (further detailed in \secref{sec:ILP}).

\fakepar{Adaptability}
\name satisfies the runtime adaptability requirement of \cps applications by \emph{switching between multiple pre-configured operation modes}, similar to \cite{fohler1993changing}.

The beacons, sent by the host at the beginning of each round, contain the current round \id, as well as a mode \id and a trigger bit \SB used in the mode change procedure as described below.

\begin{figure}
\begin{minipage}[c]{0.7\linewidth}
\centering
\includegraphics[width=0.8\linewidth]{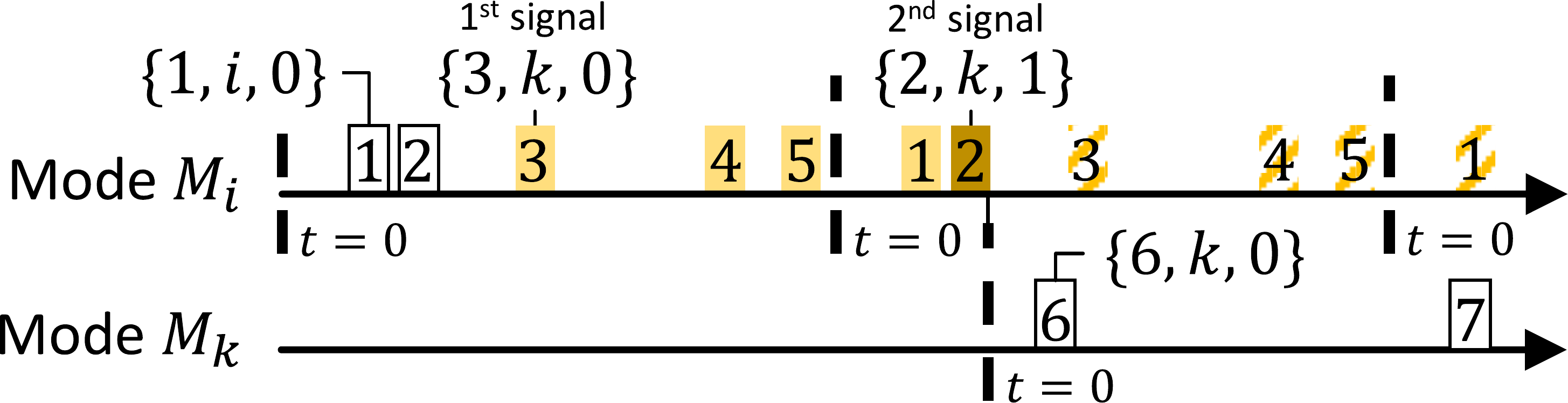}%
\end{minipage}%
\hfill
\begin{minipage}[c]{0.3\linewidth}
\centering
\begin{align*}
\beacon \; = \; \{ \; 
	&\,\text{round }\id \\[-2pt]
	&\,\text{mode }\id \\[-2pt]
	&\,\text{trigger bit }{\SB} \; \}
\end{align*}
\end{minipage}%
\caption{%
Example of mode change from \mode{i} to \mode{k}.
\capt{%
The dashed lines show the start of the mode hyperperiods.
Each numbered box represents a communication round with its \id.
The content of some of the host beacons is shown next to the corresponding rounds.
Steady-state rounds are shown in white, \eg $r_6$.
In $r_3$, the host sends the first signal to change to mode \mode{k}. During this transition phase, the rounds are lightly colored.
In the dark colored round, the host sets $\SB = 1$. This is the second signal -- directly after this round, mode \mode{k} starts executing.
Dashed rounds are not executed.
}
}
\label{fig:mode_change}
\end{figure}

A mode change happens in two phases. First, all running applications finish their execution but new application instances don't start. Afterwards, the new mode starts with its first round.
The host triggers mode changes using its beacons as illustrated in \figref{fig:mode_change}. 
In the first phase, the host sends the \ids of rounds from the current mode together with
the \id of the new mode. This allows the nodes to prepare and to not start new applications. The host statically knows when there is no application running in the system anymore. 
Then, at the next communication round, the host sends the round \id, the \id of the new mode, and set the trigger bit $\SB = 1$. The new mode starts directly after this communication round ends. 

\fakepar{Energy efficiency}	
As \name uses static scheduling, one can \emph{distribute the schedules} at deployment time to limit the protocol communication overhead at runtime, thus optimizing energy efficiency.
At deployment time, the node's task and communication schedule is loaded into its memory. It contains, for each mode, the relative starting times for the mode's communication rounds, the mode's hyperperiod, the slots allocated to node $n$ for each round, \ie pairs (slot \id; message \id), and the number of slots allocated in each round\footnote{This enables to save energy if less than $B$ slots are allocated.}.

\fakepar{Predictability and Reliability}
Thanks to the distributed schedule information, it is sufficient for any node to receive a single beacon to retrieve the overall system state -- \ie the phase of the schedule given by the round \id -- thus which message to send in which slot and when to wake up for the next communication round. 
Modes and rounds have unique \ids. Let us denote the beacon by \beacon. 
If $\beacon = \{j, k, 1\}$, then the next round is the \emph{first round scheduled in mode with $\id = k$}. Otherwise, if $\beacon = \{j, k, 0\}$ then the next round is the next one after $\id=j$ in the cyclic sequence of rounds associated to the mode $\id=k$. If round with $\id=j$ is not part of mode with $\id=k$, then no new applications are started.

If a node does not receive the beacon, it does not participate in the current communication round. This way it is guaranteed that a packet loss does not lead to message collisions%
\footnote{Assuming conflict-free schedules, see \secref{sec:ILP} for the schedule synthesis.}%
.

Altogether, those concepts guarantee that \name executes predictably and reliably.
In contrast to LWB~\cite{ferrari2012low}
, the above system design allows for fast mode switches, guarantees safe operation in terms of non-overlapping communication, combines task- and communication scheduling, and combines offline with on-line scheduling decisions.

\section{System Model and Scheduling Problem}\label{sec:model}


Each distributed \emph{application} is composed of \emph{tasks} and \emph{messages} connected by precedence constraints described by a directed acyclic graph, where vertices and edges represent tasks and messages, respectively. We denote by \app.\predG the \emph{precedence graph} of application \app.
Each application executes at a periodic interval $\app.p$, called the \emph{period}.
An application execution is completed when all tasks in \predG have been executed.
All tasks and messages in \app.\predG share the same period, $\app.p$.
Applications are subject to real-time constraints.
The application \emph{relative deadline}, denoted by $\app.d$ with $\app.d \leq \app.p$, represents the maximum tolerable \emph{end-to-end delay} to complete the execution.
In summary, an application \app is characterized by
\begin{align*}
\app \; = \; \{ \quad
	 \app.p
	\quad &\text{--\quad period [given]} \\
	 \app.d
	\quad &\text{--\quad end-to-end deadline [given]} \\
	 \app.\predG
	\quad &\text{--\quad precedence graph [given]} \quad\quad \}
\end{align*}

We denote with \taskset and \messageset the sets of tasks and messages, respectively.
A node executes at most one task at any point in time and we consider \emph{non-preemptive task scheduling}. Each task $\tau$ is mapped to a given node $\tau.map$, on which it executes within a WCET $\tau.e$%
\footnote{
Nodes are assumed capable of performing one task execution and one message transmission simultaneously, as supported by several state-of-the-art wireless \cps platforms featuring two cores, \eg~\cite{Freescale}.}
.
The \emph{task offset} $\tau.o$ represents the start of the task execution, relative to the beginning of the application execution.
A task can have an arbitrary number 
of preceding messages, \ie messages that must be received before it can start. $\tau.prec$ denotes the set of preceding message \ids.
In summary, a task $\tau$ is characterized by
\begin{align*}
\tau \; = \; \{ \quad
	 \tau.o
	\quad &\text{--\quad offset {\bf [computed]}} \\
	 \tau.\map
	\quad &\text{--\quad mapping [given]} \\
	 \tau.e
	\quad &\text{--\quad WCET [given]} \\
	 \tau.\prec
	\quad &\text{--\quad preceding message set [given]} \\
	 \tau.p
	\quad &\text{--\quad period [given, equal to \app.p]} \quad \quad \}
\end{align*}

\begin{figure}
\centering
\includegraphics[width = 0.6\linewidth]{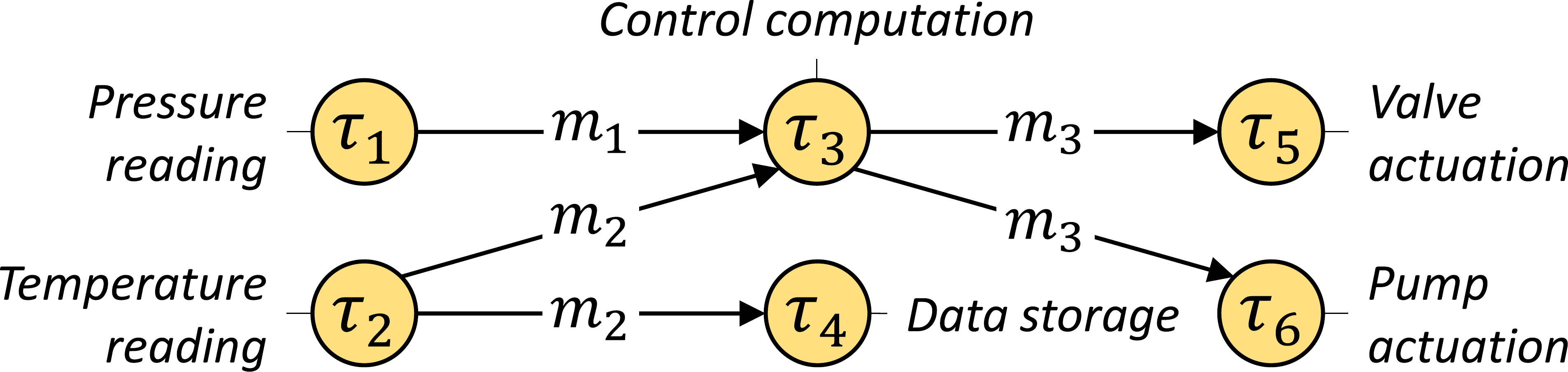}
\vspace{-0.15cm}
\caption{An example control application and its precedence graph \predG.
\capt{The execution starts with sensor readings -- either $\tau_1$ or $\tau_2$. After both have been received by the controller, actuation values are computed ($\tau_3$), multicast to the actuators ($m_3$), and applied ($\tau_5$ and $\tau_6$).
}}
\label{fig:precedence_graph}
\vspace{-0.5cm}
\end{figure}

%
Every message $m$ has at least one preceding task, \ie tasks that need to finish before the message can be transmitted. The set of preceding task \ids is denoted by $m.prec$. All preceding tasks must be mapped to the same node.
The \emph{message offset} $m.o$, relative to the beginning of the application execution, represents the earliest time message the $m$ can be allocated to a round for transmission, \ie after all preceding tasks are completed.
The \emph{message deadline} $m.d$, relative to the message offset, represents the latest time when the message transmission must be completed, \ie the earliest offset of successor tasks.
All messages have the same payload. 
In summary, a message $m$ is characterized by
\begin{align*}
m \; = \; \{ \quad
	 m.o
	\quad &\text{--\quad offset {\bf [computed]}} \\
	 m.d
	\quad &\text{--\quad deadline {\bf [computed]}} \\
	 m.\prec
	\quad &\text{--\quad preceding task set [given]} \\
	 m.p
	\quad &\text{--\quad period [given, equal to \app.p]} \quad \quad \}
\end{align*}
%
%
\noindent
Within one application \app, each task is unique.
Messages are not necessarily unique, \ie multiple edges of \app.\predG can be labeled with the same message $m$, which captures the case of multicast/broadcast.
If a task $\tau$ or a message $m$ belongs to two different applications \appi and \appj, the two applications have the same period, \ie $\appi.p = \appj.p$. \figref{fig:precedence_graph} shows a control application and its precedence graph as an example.

\emph{Operation modes} represent mutually exclusive phases of the system, \eg \emph{boostrapping}, \emph{normal}, and \emph{emergency} modes, each executing a specific schedule. A mode $\mode{}$ is a set of applications \app that are concurrently executed, that is
\begin{align*}
\mode{} \; = \; \{ \quad
	 \appi\, , \;
	 \appj\, , \; \ldots
	\quad &\text{--\quad applications to execute [given]}  \quad \}
\end{align*}
The mode \emph{hyperperiod} is the least common multiple of the mode's applications.
In this paper, we assume \emph{no intersection between the modes}%
\footnote{Supporting more general models induces additional constraints, studied \eg in~\cite{fohler1993changing}. Tackling those is beyond the scope of this paper.}%
,
\ie $\modei \cap \modej = \emptyset$ if $i \neq j$.

Finally, the schedule of a mode \mode{} contains $R_{\mode{}}$ \emph{communication rounds} $r$.
Rounds are \emph{atomic}, that is, they cannot be interrupted.
Each round is composed of (up to) $B$ slots, each allocated to a unique message $m$. This results in a maximum round length \Tround.
The \emph{round starting time} $r.t$ is the start of the round relative to the beginning of the mode hyperperiod.
The \emph{allocation vector} $r.[B]$ is a vector of size $B$ containing the \ids of the messages allocated to the slots. $r.B_s$ denotes the allocation of the $s$-th slot.
In summary, a round $r$ is characterized by
\begin{align*}
r \; = \; \{ \quad
	 r.t
	\quad &\text{--\quad starting time {\bf [computed]}} \\
	 r.[B]
	\quad &\text{--\quad allocation vector {\bf [computed]}}  \quad \}
\end{align*}


We consider that all modes, applications, task mappings and WCETs are given. For a given mode \mode{}, the remaining variables define the mode schedule, denoted by \sched{\mode{}}:
\begin{align*}
&\sched{\mode{}} \, = \,
	\left\lbrace
	\begin{tabular}{c|l}
	$\tau.o, \, m.o, \, m.d$
	&
	$\app \in \mode{}, \;
	(\tau,m) \in \app.\predG$
	\\
	$r_k.t, \, r_k.[B]$
	&
	$k \in [1, \, R_{\mode{}}]$
	\end{tabular}	 
	\right\rbrace
\end{align*}

\fakepar{Scheduling Problem} 
Design a scheduling scheme that, given a mode \mode{}, returns an optimized mode schedule \sched{\mode{}} s.t.\linebreak
	\emph{(i)} application executions always meet their deadline, and \\
	\emph{(ii)} the number $R_\mode{}$ of communication rounds is minimized.

\section{\name Scheduling}\label{sec:ILP}

As introduced in \secref{sec:implementation}, \name statically synthesizes the schedule of all tasks, messages, and communication rounds to meet real-time constraints, minimize end-to-end latency, and minimize the energy consumed for communication. 
This section presents the ILP formulation used and how to solve the scheduling problem, known to be NP-hard~\cite{jeffay1991NPnonpreemptive}, efficiently.

The schedule of a mode \mode{} is computed for one hyperperiod, after which it repeat itself. 
To minimize the number of rounds used while handling complexity, we solve the problem sequentially, as described in Alg.~\ref{alg:outerlayer}. 
Each ILP formulation considers a fixed number of rounds $R_{\mode{}}$ to be scheduled, starting with $R_{\mode{}}=0$. The number of rounds is incremented until a feasible solution is found, or until the maximum number of rounds $R_{max}$ -- the number of rounds that ``fit'' into one hyperperiod -- is reached.
Thus, Alg.~\ref{alg:outerlayer} guarantees by construction that if the problem is feasible, the synthesized schedule is optimal in terms of number of rounds used.
The end-to-end latency is minimized by setting the sum of all application's latency as objective function.

\begin{algorithm}
\begin{algorithmic}
\footnotesize
\Require 
	mode \mode{}, 
	applications $\app \in \mode{}$,
	task mappings $\tau.\map$ and WCETs $\tau.e$, 
	round duration $\Tround$
	\; -- \;
\textbf{Output:}
	\sched{M}

\State $LCM \gets$ \textit{hyperperiod}(\mode{})
\State $R_{max} = floor(LCM/\Tround)$

\State $R_{\mode{}} = 0$

\While{$R_{\mode{}} \leq R_{max}$} 
	\State formulate the ILP for mode \mode{} using $R_{\mode{}}$ rounds
	\State [ \sched{M}, \textit{feasible} ] = \textit{solve}( ILP )
	\If {\textit{feasible}}
		\Return \sched{M}
	\EndIf
	\State $R_{\mode{}} \gets R_{\mode{}}+1$
\EndWhile
\State \Return 'Problem infeasible'
\end{algorithmic}
\caption{\small Pseudo-code of the schedule synthesis}
\label{alg:outerlayer}
\end{algorithm}

The ILP formulation contains a set of classical scheduling constraints.
%
%
	\emph{Precedence constraints} between tasks and messages must be respected;
%
	\emph{end-to-end deadlines} must be satisfied;
%
	nodes process \emph{at most one task} simultaneously;
%
	rounds must \emph{not overlap};
%
%
	rounds cannot be allocated more messages than the maximal number of slots available.
%
Those constraints can be easily formulated using our system model (see Appendix for the full formulation).
However, one must also guarantee that the allocation of messages to rounds is valid, \ie  
\emph{every message must be served in a round that starts after its release time \emph{\textbf{(C1)}} and finishes before its deadline \emph{\textbf{(C2)}}}. This creates a non-linear coupling between the variables and makes the problem not trivial. 
\emph{This is the key difference with the existing approaches for wired architectures, like}~\cite{craciunas2016coscheduling}.

\begin{figure}
\centering
\includegraphics[width=0.6\linewidth]{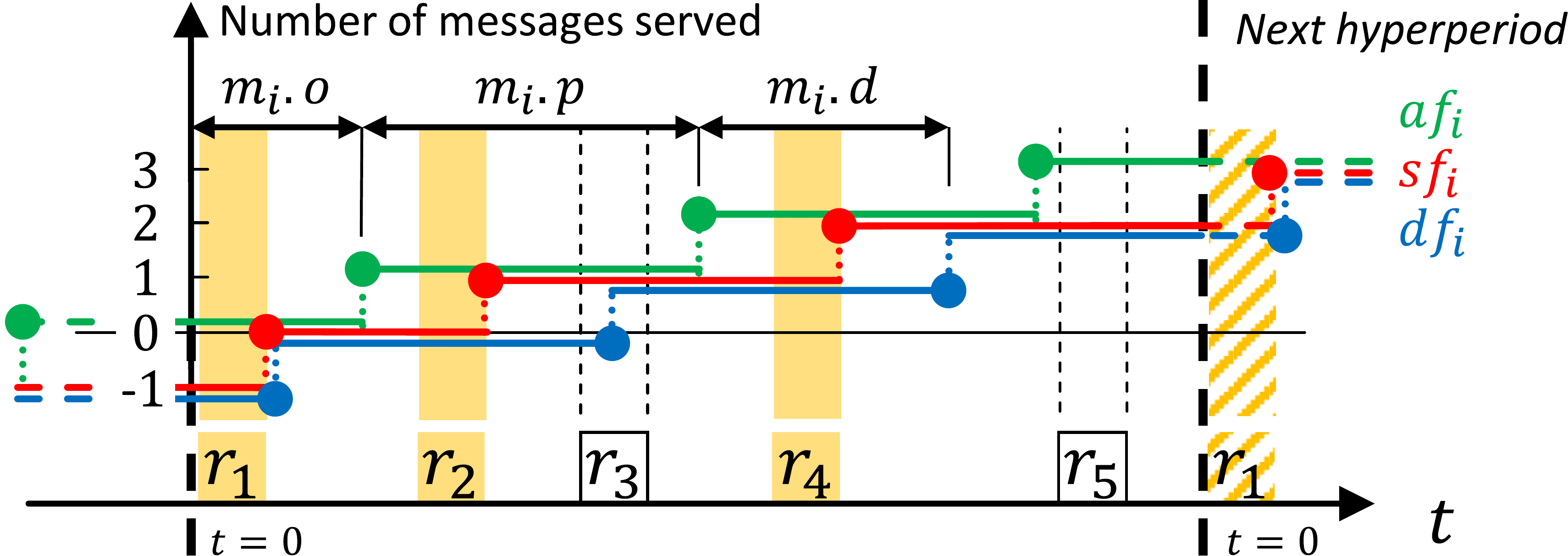}
\caption{Representation of arrival, demand, and service functions of message $m_i$. 
The lower part shows the five round, $r_1$ to $r_5$, scheduled for the hyperperiod. 
\capt{%
$m_i$ is allocated a slot in the colored rounds, \ie $r_1$, $r_2$, and $r_4$. 
The allocation of $m_i$ to $r_3$ instead of $r_2$ would be invalid, as $r_3$ does not finish before the message deadline, \ie it violates \textbf{(C4.2)}.
However, the allocation of $m_i$ to $r_5$ instead of $r_1$ would be valid and result in $r_0.B_i = 0$.}
}
\label{fig:afdfsf}
\end{figure}

To address this challenge, we first formulate the constraints \textbf{(C1)} and \textbf{(C2)} using \emph{arrival}, \emph{demand}, and \emph{service} functions, \af \df and \sf, using network calculus~\cite{leboudec2001network}. 
Those functions count the number of message instances released, with passed deadlines, and served since the beginning of the hyperperiod, respectively. 
Those three functions are illustrated in \figref{fig:afdfsf}.
It must hold that
\begin{flalign}
\label{eq:df<sf<af}
&\forall\, m_i \in \messageset, \;\forall\, t, 
&&\df_i(t) \leq \sf_i(t) \leq \af_i(t)
&&\\
\label{eq:af_def}
&\text{with},
&&\af_i: \; t \;
	\longmapsto \; \left \lfloor{\frac{t-m_i.o}{m_i.p}}\right \rfloor 	+ 1
	&&\\
\label{eq:df_def}
&\text{and},
&&\df_i: \; t \;
	\longmapsto \; \left \lceil{\frac{t-m_i.o-m_i.d}{m_i.p}}\right \rceil 
	&&
\end{flalign}

\noindent
However, as the service function stays constant between the rounds, we can formulate \textbf{(C1)} and \textbf{(C2)} as follows\\
$\forall\, m_i \in \messageset, \; \forall\, j \in [1 .. R_{\mode{}}], $
\begin{flalign}
\label{eq:af_const}
&\textbf{(C1)} \quad  : \quad 
	&\sf_i(r_j.t + \Tround) \, &\leq \, \af_i(r_j.t)
	&& 
\\
\label{eq:df_const}
&\textbf{(C2)} \quad  : \quad
	&\sf_i(r_j.t)  \, &\geq \, \df_i(r_j.t + \Tround)
	&& 
\end{flalign}

The arrival and demand functions are step functions. They cannot be used directly in an ILP formulation, however
\begin{flalign}
\label{eq:af=k}
&\forall \; k \in \mathbb{N}, \quad
&&\af_i(t) = k 
	\quad \Leftrightarrow \quad
	0 \, \leq \, t - m_i.o - (k-1)m_i.p \,<\, m_i.p &&\\
&\text{and} 
&&\df_i(t) = k 
\label{eq:df=k}
	\quad \Leftrightarrow \quad
	0 \, < \, t - m_i.o - m_i.d - (k-1)m_i.p \,\leq\, m_i.p &&
\end{flalign}
For each message $m_i\in \messageset$ and each round $r_j$, $j \in [1..R_{\mode{}}]$, we introduce two integer variables $k^a_{ij}$ and $k^d_{ij}$ that we constraint to take the values of \af and \df at the time points of interest, \ie $r_j.t$ and $r_j.t + \Tround$ respectively. That is,
\begin{align}
\label{eq:ka} 
0 \, \leq \, r_j.t  
	&-m_i.o - (k^a_{ij}-1)m_i.p \,<\, m_i.p\\
\label{eq:kd} 
0 \, < \, r_j.t  
	&+\Tround - m_i.o - m_i.d - (k^d_{ij}-1)m_i.p \,\leq\, m_i.p\\
\notag
\text{Thus,} \hspace{15pt} &\eqref{eq:ka} \quad \Leftrightarrow  \quad 
	 \af_i(r_j.t) = k^a_{ij} \\
\notag
	&\eqref{eq:kd} \quad \Leftrightarrow \quad 
	\df_i(r_j.t + \Tround) = k^d_{ij}
\end{align}

Finally, we must express the service function \sf, which counts the number of message instances served \emph{at the end} of each round. 
Remember that $r_k.B_s$ denotes the allocation of the $s$-{th} slot of $r_k$. 
For any time $t \in \; [ \; r_{j} + \Tround \, ; \,  r_{j+1} + \Tround \; [$, the number of instances of message $m_i$ served is
\begin{align*}
	\sum_{\substack{k = 1}}^{j} \;\;
	\sum_{\substack{s = 1}}^{B}
	 \; r_k.B_s
	 \quad s.t. \; B_s = i
\end{align*}

\noindent
It may be that $m.o + m.d > m.p$, resulting in $\df(0)=-1$ (see \eqref{eq:df_def}), like \eg in \figref{fig:afdfsf}. This ``means'' that a message released at the each of one hyperperiod will have its deadline in the \emph{next} hyperperiod. 
To consider this situation, we introduce, for each message $m_i$, a variable $r_0.B_i$ set to the number of such ``leftover'' message instances at $t=0$. Finally, for each message $m_i \in \messageset$, and  $t \in \; [ \; r_{j} + \Tround \, ; \,  r_{j+1} + \Tround \; [$,
\begin{flalign}
\label{eq:sf_def}
\sf_i: \; t \;
	&\longmapsto \;
	\sum_{\substack{k = 1 \\[2pt]s.t. \; r_k + \Tround \, < \, t}}^{j}\;\;
	\sum_{\substack{s = 1 \\[2pt]s.t. \; B_s = i}}^{B}
	 r_k.B_s - r_0.B_i 
\end{flalign}

\noindent
Ultimately, \textbf{(C1)} and \textbf{(C2)} can be formulated as ILP constraints using \eqref{eq:ka}, \eqref{eq:kd}, and the following two equations:
\begin{align}
&
\text{Eq.\,}\eqref{eq:af_const}\quad	 \Leftrightarrow
	\quad	
	\sum_{k = 1}^j
	\sum_{\substack{s = 1 \\[2pt]s.t. \; B_s = i}}^{B}
	 \; r_k.B_s - r_0.B_i \; \leq\;  k^a_{ij}
\\
&
\text{Eq.\,}\eqref{eq:df_const}\quad	 \Leftrightarrow
	\quad	
	\sum_{k = 1}^{j-1}
	\sum_{\substack{s = 1 \\[2pt]s.t. \; B_s = i}}^{B} \; r_k.B_s - r_0.B_i 
	\; \geq\; 
	k^d_{ij}
\end{align}


\section{Evaluation}\label{sec:evaluation}
%

We evaluate the performance of \name regarding two criteria: the minimal achievable latency for an application, and the energy savings from the use of communication rounds.

Let $\app.\delta$ denote the latency of an application \app. It represents the delay for a complete execution of \app, \ie the completion of all tasks in \app.\predG. Let $\app.c$ be a \emph{chain} in \app.\predG. A chain is a path of \app.\predG starting and ending with a task without predecessor and successor, respectively. For example, $(\tau_2, m_2, \tau_4)$ is a path of \predG in \figref{fig:precedence_graph}. 
The minimum achievable latency for a single message in \name is \Tround. Thus $\app.\delta$ is lower-bounded by
\begin{align}
\label{eq:min_deadline}
\app.\delta \; 
	& \geq \; \max_{\app.c \,\in\, \app.\predG} 
		\left( 
			\; \sum_{\tau \, \in \, \app.c} \tau.e \,+\, \sum_{m \, \in \, \app.c} \Tround \;
		\right)
\end{align}

As discussed in \secref{sec:relWork}, to the best of our knowledge,~\cite{jacob2016rtss} is the only approach that provides timing guarantees similar to \name.
However, the best possible guarantee for a single message is of the order of $2*\Tround$~\cite{jacob2016rtss}, due to the loose coupling between the task and message schedules.
Hence, our approach improves the message latency by (at least) a factor 2.

\begin{figure}
\centering
\includegraphics[width=0.6\linewidth]{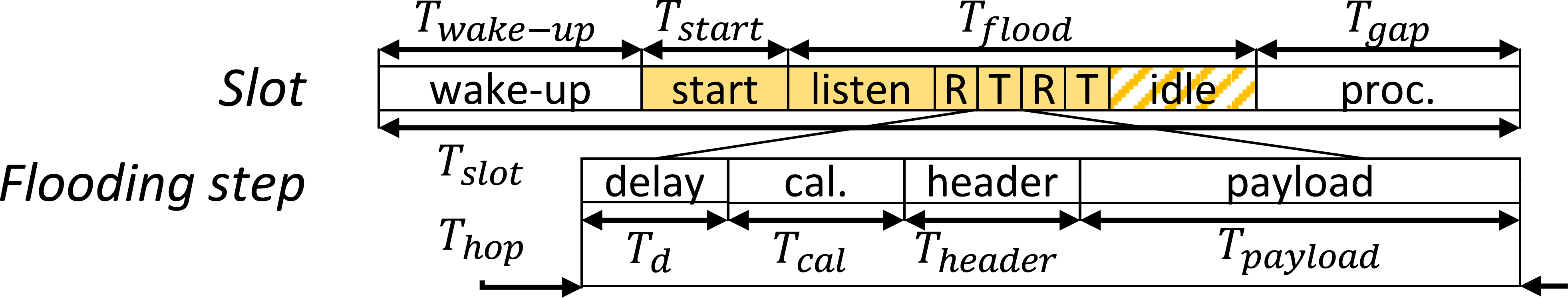}
\caption{%
Break-down of a communication round.
\capt{%
At the slot level, the colored boxes identify phases where the radio is on.
In the ``idle'' phase, the radio is turned off in practice, but this idle time of each node depends on its distance to the initiator.
To evaluate the energy benefits of rounds (see \figref{fig:energy_ratio}), we assume the radio stays on for the whole \Tglossy time~\eqref{eq:Ton}.
}
}
\label{fig:Tslot}
\end{figure}

To minimize the application latency, one can minimize the round length \Tround, down to a certain limit that we investigate now. 
A round is composed of $(B+1)$ slots (see \figref{fig:overview}) taking each a time $\Tslot(l)$ to complete, where $l$ the payload size in Bytes.
The composition of each slot is detailed in \figref{fig:Tslot}.
First, all nodes wake up (it takes \Twakeup) and switch on their radio (\Tstart).
Then the message flood starts (see~\figref{fig:overview}(b)). We denote by \Thop the time required for one protocol step, \ie a one-hop transmission. The total length of the flood is
\begin{align}
\Tglossy = (H+2N-1)\Thop 
\end{align}

\noindent
with $H$ the network diameter and $N$ the number of times each node transmits each packet\footnote{Glossy achieves more than 99.9\% packet reception rate using $N = 2$~\cite{ferrari11}.}.
\Thop is itself divided into
\begin{align}
\Thop = \Td + \Tcal + \Theader + \Tpayload
\end{align}

\noindent
where \Td is a radio delay, and \Tcal, \Theader and \Tpayload are the transmission times of the clock calibration message, the protocol header and the message payload, respectively.
With a bit rate of \Rbit, the transmission of $l$ Bytes takes time
\begin{align}
T(l) = 8l/\Rbit
\end{align}

\noindent
Once the message flood is completed, some time \Tgap is necessary to process the received packet.
We split \Tslot into \Ton and \Toff, the time spent with radio on and off. $\Tslot(l) = \Toff + \Ton(l)$ with
\begin{align}
\Toff &\;=\;  
	\Twakeup + \Tgap \\
\label{eq:Ton}
\Ton(l) &\;=\;  
	\Tstart + (H+2N-1) * \left( \Td + 8(\Lcal + \Lheader + l)/\Rbit \right) \\
\text{and \qquad }\Tround(l) 
	&\;=\; 
	\Tslot(\Lbeacon) + B*\Tslot(l)
\end{align}

\begin{figure}
\centering
\includegraphics[scale=0.75]{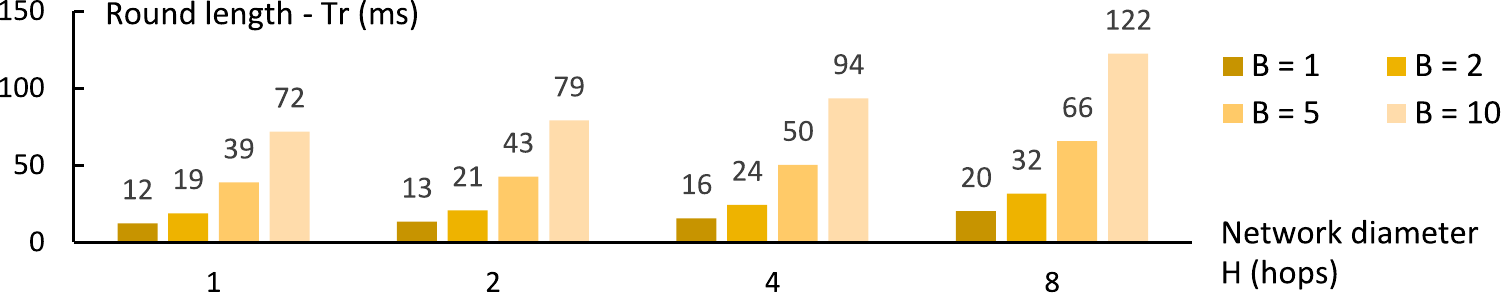}
\caption{%
Sample values of the round length \Tround
for different network diameters and numbers of slot per round. %
\capt{%
Payload is $l=10$ B and $N = 2$.}
}
\label{fig:Tround_values}
\end{figure}

\noindent
With our implementation, a beacon size of \Lbeacon = 3 Bytes is sufficient. 
We use the values from Table~\ref{table:CC430values} to derive \Tround as function of the network diameter $H$ and the number of slots per round $B$ (see \figref{fig:Tround_values}).
It shows \eg a minimum message latency of 50\ms in a 4-hop network using 5-slot rounds.
The complete numerical model is available as supplementary material.

We now consider the benefits of using rounds in terms of energy, using the radio-on time as metric%
.
Beacons are necessary to reliably prevent message collisions (see \secref{sec:implementation}). 
In a design \emph{without} round, each message transmission is preceded by its own beacon, such that the transmission time for $B$ messages of size $l$, denoted $\Tworound(l)$, takes
\begin{align}
\Tworound(l) =  B*( \, \Tslot(\Lbeacon) + \Tslot(l) \, )
\end{align}

\begin{figure}
\centering
\includegraphics[scale=0.75]{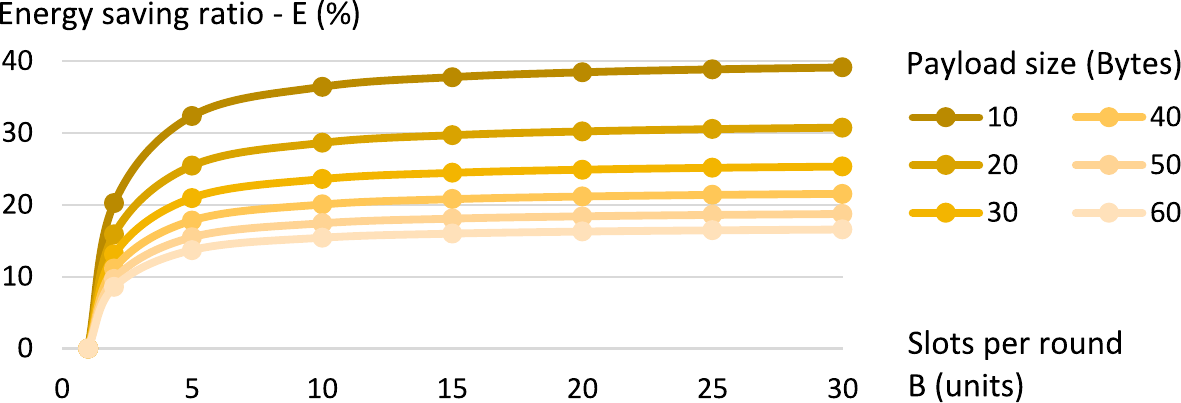}
\caption{Relative radio-on time benefit of using rounds compared to single messages. $H = 4$, $N = 2$.
\capt{As each round requires only one beacon from the host, the benefit of using rounds grows with the number of number of slots per round (X-axis). Conversely, those savings become less significant as the payload size increases (lighter colors).}}
\label{fig:energy_ratio}
\end{figure}

\noindent
Using \eqref{eq:Ton}, we compute the relative energy saving $E = (\Tworoundon - \Troundon)/\Tworoundon$ as a function of the payload size $l$ and the number of slots $B$. As shown in \figref{fig:energy_ratio}, 5-slot rounds already induce 33\% energy savings for 10 Bytes of payload.

Will a physical implementation yield comparable energy and delay results? 
Models of Glossy and LWB show very close correlations to the measured performance of the physical system~\cite{ferrari11,ferrari2012low}.
Based on this fact and the high similarity with our model, we can expect that yes, an implementation of \name will match the results of the performance modeling with high accuracy,
thus validating \name's design for \cps applications.

\begin{table}[t]
\centering
\begin{tabular}{@{}c|c|c|c|c|c|c@{}}
\Twakeup
	& 
\Tstart
	& 
\Td
	& 
\Lcal
	& 
\Lheader
	& 
\Tgap
	& 
\Rbit
	\\[3pt]
	\hline
	750 \us
	& 
	164 \us
	& 
	68 \us
	& 
	3 B
	& 
	6 B
	& 
	3 \ms
	& 
	250 kbps
\end{tabular}
\vspace{5pt}
\caption{Constants of a publicly available Glossy implementation~\cite{LWBCodeCC430}}
\label{table:CC430values}
\end{table}

\section{Related Work}\label{sec:relWork}

Various high reliability protocols have been proposed for low-power multi-hop wireless network, like TSCH~\cite{TSCH}, WirelessHART~\cite{hart} or LWB~\cite{ferrari2012low}. 
Despite their respective benefits, all those protocols consider only network resources. They do not take into account the scheduling of distributed tasks on the computation resources, and therefore they hardly support end-to-end deadlines as commonly required for \cps applications~\cite{akerberg11}.
In \cite{jacob2016rtss}, a protocol that provides such end-to-end guarantees is proposed, but it couples tasks and messages as loosely as possible, aiming for efficient support of sporadic or event-triggered applications.
This results in high worst-case latency and is thus not suitable for demanding \cps applications~\cite{akerberg11}.
%
%
This observation points toward a fully time-triggered system where tasks and messages are co-scheduled.

In the wired domain, much work has been done on time-triggered architecture, like TTP~\cite{kopetz93}, the static-segment of FlexRay~\cite{FlexR}, or TTEthernet~\cite{kopetz05}.
Many recent works use SMT- of ILP-based methods to synthesize and/or analyze static (co-)schedules for those architectures~\cite{steiner2010evaluation,craciunas2016coscheduling,ashjaei2017end2endReservation,tamas2012synthesis,zhang14}.
However, these approaches assume that a message can be scheduled at anytime. While being a perfectly valid hypothesis for a wired system, this assumption is not compatible with the use of communication rounds in a wireless setting.
%
%
%
%
%
%
%
%
%

\section{Conclusions}\label{sec:conclusion}


This paper presented Time-Triggered-Wireless (\name), a distributed low-power wireless system design, motivated by the need for a wireless solution providing low latency, energy efficiency and high reliability for time critical applications, \eg industrial closed-loop control systems.
This paper presented the design concepts that enable \name to meet those requirements. 
The performance evaluation of \name, based on low-level system models, shows a reduction of communication latency by a factor 2x and of energy consumption by 33-40\% compared to the closest related work~\cite{jacob2016rtss}.
This validates the suitability of \name for wireless \cps applications and opens the way for implementation and real-world experience with industry partners.

\section*{Acknowledgments}
This work is supported by Nano-Tera.ch, with Swiss Confederation financing, and by the German Research Foundation (DFG) within cfaed and SPP 1914, project EcoCPS.

\pagebreak
\bibliographystyle{IEEEtran}
\bibliography{IEEEabrv,refs_networking,refs_staticSched}

\pagebreak
\appendix


This appendix presents the complete description of the ILP formulation, used to synthesize the schedule \sched{\mode{}} of a given operation mode \mode{}, \ie 
\begin{align*}
&\sched{\mode{}} \; = \;
	\left\lbrace
	\begin{tabular}{c@{\quad}|@{\quad}l}
	$\tau.o, \, m.o, \, m.d$
	&
	$\app \in \mode{}, \;
	(\tau,m) \in \app.\predG$
	\\
	$r_k.t, \, r_k.[B]$
	&
	$k \in [1, \, R_{\mode{}}]$
	\end{tabular}	 
	\right\rbrace
\end{align*}

The system model is described in \secref{sec:model}. All the variables and parameters involved in the ILP formulation used to solve the single-mode schedule synthesis problem are listed at the end in Table~\ref{tab:ILPvar} for reference.
This appendix details all the formulated constraints and how they are implemented in practice. The last subsection details the formulation of the objective function used to minimize the end-to-end application latency.

The ILP formulation to be solved by Alg.~\ref{alg:outerlayer} (see \secref{sec:model}) contains the following constraints, which can be classified into four categories.

\fakepar{1. Application constraints}
		\begin{description}
			\item[(C1.1)] Precedence constraints between tasks and messages must be respected.
			\item[(C1.2)] End-to-end deadlines must be satisfied.
		\end{description}
			
\fakepar{2. Round constraints}	
		\begin{description}
			\item[(C2.1)] The rounds must not be overlapping.
			\item[(C2.2)] The time interval between two rounds is upper-bounded.
		\end{description}
	
\fakepar{3. Validity of the tasks mapping}	
		\begin{description}
			\item[(C3)] The same node cannot process more than one task simultaneously.
		\end{description}
	
\fakepar{4. Validity of the messages allocation}	
		\begin{description}
			\item[(C4.1)] Every message must be served 
			after its release time.
			\item[(C4.2)] Every message must be served 
			before its deadline.
			\item[(C4.3)] A round cannot be allocated more messages than the number of slots available in one round.
			\item[(C4.4)] Within one hyperperiod, the same number of messages are released and served.
		\end{description} 

The validity of the message allocation constraints, in particular constraints \textbf{(C4.1)} and \textbf{(C4.2)}, induce a non-linear coupling between the message and round variables. This peculiar aspect makes the scheduling problem non-trivial and prevents the use of conventional ILP-based schedule synthesis approaches reported thus far in the literature.
Our approach to solve this problem is detailed in \secref{sec:message_alloc}.

\subsection*{Application constraints}\label{sec;application}
\begin{description}
	\item[(C1.1)]Precedence constraints between tasks and messages must be respected.
\end{description}
For any application \app, let $\app.c$ be a \emph{chain} in \app.\predG. A chain is a path of \app.\predG starting and ending with a task without predecessor and successor, respectively. $\app.c(\,\first\,)$ and $\app.c(\,\last\,)$ denote the first and last task of the chain \app.c.  $\sigma_{i,j}$ is a binary variable accounting for the cases where successive task $\tau_i$ and message $m_j$ (resp. a message $m_i$ and task $\tau_j$) start during different application period ($\sigma_{i,j} = 1$) or not ($\sigma_{i,j} = 0$)
\begin{align}		
\intertext{%
$	\forall\, \app, \; 
	\forall\, \app.c \in \app.\predG, \;  
	\forall\, \tau_j \in (\app.c \setminus \app.c(\,\last)),\;  
	\forall\, m_i \in \tau_j.prec,
	$}
		m_i.o + m_i.d
		&\;\leq\; \app.p * \sigma_{i,j} + \tau_j.o\\
\intertext{%
$	\forall\, \app, \;  
	\forall\, \app.c \in \app.\predG, \;  
	\forall\, m_j \in \app.c,\;  
	\forall\, \tau_i \in m_j.prec,
	$}
		\tau_i.o + \tau_i.e 
		&\;\leq\; \app.p * \sigma_{i,j} + m_j.o
\end{align}

\begin{description}
	\item[(C1.2)] End-to-end deadlines must be satisfied.
\end{description}
\begin{align}
\intertext{%
$	\forall\, \app, \; 
	\forall\, \app.c \in \app.\predG, \;   
		\tau_{\first} = \app.c(\,\first\,), \; 
		\tau_{\last} = \app.c(\,\last\,), \;
		C = |\app.c|, $}
& \tau_{\last}.o + \tau_{\last}.e - \tau_{\first}.o + \sum_{j=1}^{C-1} \app.p*\sigma_{i,j}  \;\leq\; \app.d
\end{align}

\subsection*{Round constraints}\label{sec:rounds}
\begin{description}
	\item[(C2.1)]The rounds must not be overlapping.
\end{description}
\begin{flalign}
&\forall j \in [1..R_{\mode{}}-1],
& & r_j.t + \Tround \; \leq \; r_{j+1}.t
&
\end{flalign}

\begin{description}
	\item[(C2.2)]The time interval between two rounds is upper-bounded.
\end{description}
\begin{flalign}
&\forall j \in [1..R_{\mode{}}-1],
&
&r_{j+1}.t - r_j.t  \; \leq \;  T_{max}
&
&
\end{flalign}

\subsection*{Validity of the task mappings}\label{sec:task_mapping}
\begin{description}
	\item[(C3)]The same node cannot process more than one task simultaneously.
\end{description}
\begin{align}
\intertext{%
$	\forall\, \tau_i, \tau_j, \; \tau_i.map == \tau_j.map, \;
	\forall\, k_i \in [1..LCM/\tau_i.p], \; \forall\, k_j \in [1..LCM/\tau_j.p]$}
\label{eq:t1beforet2}
&	\tau_i.o + \tau_i.e + \tau_i.p*k_i \; \leq \; \tau_j.o + \tau_j.p*k_j \\
\label{eq:t2beforet1}
\texttt{or} \quad 
&	\tau_j.o + \tau_j.e + \tau_j.p*k_j \; \leq \; \tau_i.o + \tau_i.p*k_i 
\end{align}
However, an ILP formulation cannot directly support that only one-out-of-two constraints must be satisfied. To resolve this, we use a classical trick using $(i)$ a binary variable $\lambda$ representing which of the two constraints \eqref{eq:t1beforet2} or \eqref{eq:t2beforet1} must be enforced and $(ii)$ a ``big'' time constant $M$, used to satisfied the other constraint by default, \ie regardless of the values of the variables. For example, $M = 10*LCM$.
\begin{align}
\label{eq:t1beforet2_impl}
\tau_i.o + \tau_i.e + \tau_i.p*k_i
	&\; \leq \; \tau_j.o + \tau_j.p*k_j + M\!M * (1 - \lambda_{i,j}^{k_i,k_j}) \\
\label{eq:t2beforet1_impl}
\tau_j.o + \tau_j.e + \tau_j.p*k_j 
	&\; \leq \; \tau_i.o + \tau_i.p*k_i + M\!M * \lambda_{i,j}^{k_i,k_j}
\end{align}
With this implementation of the constraints, it follows that
\begin{align*}
&	\lambda_{i,j}^{k_i,k_j} = 1
	\quad \Leftrightarrow \quad
		\eqref{eq:t1beforet2_impl} \equiv \eqref{eq:t1beforet2}
		\;\wedge\;
		\eqref{eq:t2beforet1_impl} \text{ is always satisfied.}
		\\
&	\lambda_{i,j}^{k_i,k_j} = 0
	\quad \Leftrightarrow \quad
		\eqref{eq:t2beforet1_impl} \equiv \eqref{eq:t2beforet1}
		\;\wedge\;
		\eqref{eq:t1beforet2_impl} \text{ is always satisfied.} 
\end{align*}

\subsection*{Validity of the messages allocation}\label{sec:message_alloc}

As mentioned earlier, the validity of the message allocation constraints, in particular constraints \textbf{(C4.1)} and \textbf{(C4.2)}, induce a non-linear coupling between the message and round variables. This peculiar aspect makes the scheduling problem non-trivial and prevents the use of conventional ILP-based schedule synthesis approaches reported thus far in the literature.
The following subsection repeats and deepens the arguments from \secref{sec:ILP}.

\begin{figure}[b]
\centering
\includegraphics[width=0.6\linewidth]{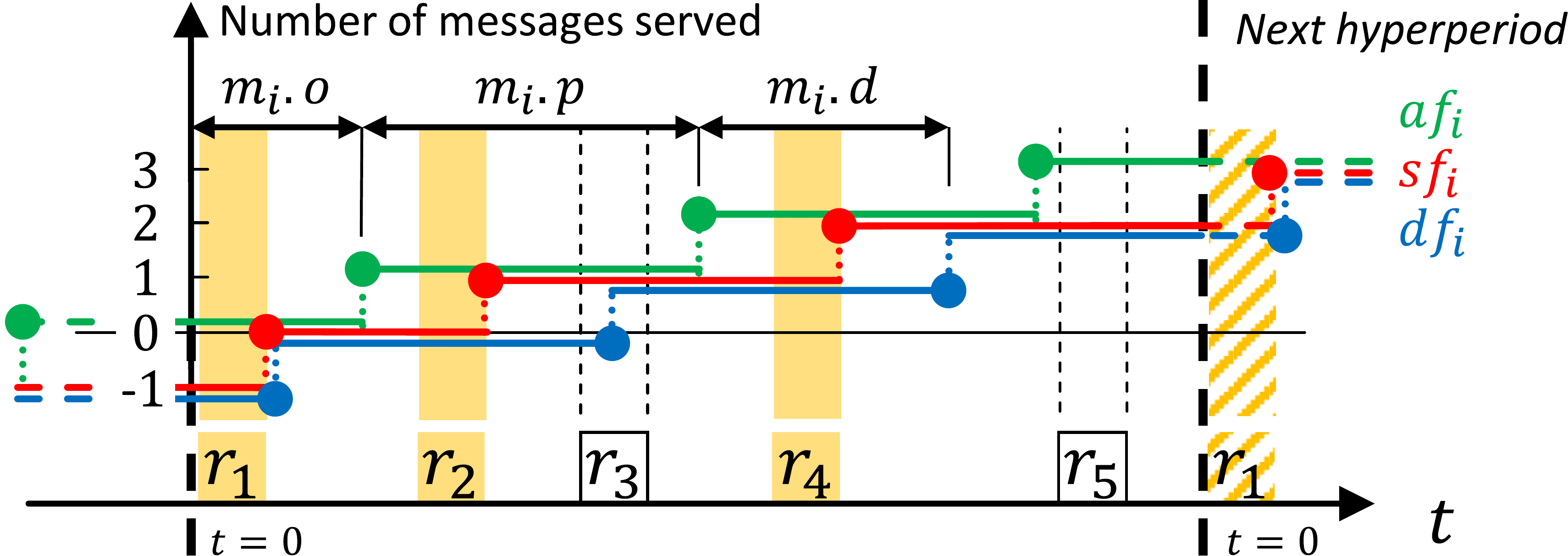}
\caption{Representation of arrival, demand, and service functions of message $m_i$. 
The lower part shows the five round, $r_1$ to $r_5$, scheduled for the hyperperiod. 
\capt{%
$m_i$ is allocated a slot in the colored rounds, \ie $r_1$, $r_2$, and $r_4$. 
The allocation of $m_i$ to $r_3$ instead of $r_2$ would be invalid, as $r_3$ does not finish before the message deadline, \ie it violates \textbf{(C4.2)}.
However, the allocation of $m_i$ to $r_5$ instead of $r_1$ would be valid and result in $r_0.B_i = 0$.}
}
\label{fig:afdfsf2}
\end{figure}

To address problem of variable coupling, we first formulate the constraints \textbf{(C4.1)} and \textbf{(C4.2)} using \emph{arrival}, \emph{demand}, and \emph{service} functions, \af \df and \sf, using network calculus%
. Those functions count the number of message instances released, served, and due since the beginning of the hyperperiod, respectively. 
Those three functions are illustrated in \figref{fig:afdfsf2}.
It must hold that
\begin{flalign}
\label{eq:df<sf<af2}
&\forall\, m_i \in \messageset, \;\forall\, t, 
&&\df_i(t) \leq \sf_i(t) \leq \af_i(t)
&&\\
\label{eq:af_def2}
&\text{with},
&&\af_i: \; t \;
	\longmapsto \; \left \lfloor{\frac{t-m_i.o}{m_i.p}}\right \rfloor 	+ 1
	&&\\
\label{eq:df_def2}
&\text{and},
&&\df_i: \; t \;
	\longmapsto \; \left \lceil{\frac{t-m_i.o-m_i.d}{m_i.p}}\right \rceil 
	&&
\end{flalign}

\noindent
However, as the service function stays constant between the rounds, we can formulate \textbf{(C4.1)} and \textbf{(C4.2)} as follows\\
$\forall\, m_i \in \messageset, \; \forall\, j \in [1 .. R_{\mode{}}], $
\begin{flalign}
\label{eq:af_const2}
&\textbf{(C4.1)}  : \quad 
	&\sf_i(r_j.t + \Tround) \, &\leq \, \af_i(r_j.t)&&
\\
\label{eq:df_const2}
&\textbf{(C4.2)}  : \quad
	&\sf_i(r_j.t)  \, &\geq \, \df_i(r_j.t + \Tround)&&
\end{flalign}

\noindent
The arrival and demand functions are step functions. They cannot be used directly in an ILP formulation, however
\begin{flalign}
\label{eq:af=k2}
&\forall \; k \in \mathbb{N}, \quad
&&\af_i(t) = k 
	\quad \Leftrightarrow \quad
	0 \, \leq \, t - m_i.o - (k-1)m_i.p \,<\, m_i.p &&\\
&\text{and} 
&&\df_i(t) = k 
\label{eq:df=k2}
	\quad \Leftrightarrow \quad
	0 \, < \, t - m_i.o - m_i.d - (k-1)m_i.p \,\leq\, m_i.p &&
\end{flalign}
For each message $m_i\in \messageset$ and each round $r_j$, $j \in [1..R_{\mode{}}]$, we introduce two integer variables $k^a_{ij}$ and $k^d_{ij}$ that we constraint to take the values of \af and \df at the time points of interest, \ie $r_j.t$ and $r_j.t + \Tround$ respectively. That is,
\begin{align}
\label{eq:ka2} 
0 \, \leq \, r_j.t  
	&-m_i.o - (k^a_{ij}-1)m_i.p \,<\, m_i.p\\
\label{eq:kd2} 
0 \, < \, r_j.t  
	&+\Tround - m_i.o - m_i.d - (k^d_{ij}-1)m_i.p \,\leq\, m_i.p\\
\notag
\text{Thus,} \hspace{15pt} &\eqref{eq:ka2} \quad \Leftrightarrow  \quad 
	 \af_i(r_j.t) = k^a_{ij} \\
\notag
	&\eqref{eq:kd2} \quad \Leftrightarrow \quad 
	\df_i(r_j.t + \Tround) = k^d_{ij}
\end{align}

Finally, we must express the service function \sf, which counts the number of message instances served \emph{at the end} of each round. 
Remember that $r_k.B_s$ denotes the allocation of the $s$-{th} slot of $r_k$. 
For any time $t \in \; [ \; r_{j} + \Tround \, ; \,  r_{j+1} + \Tround \; [$, the number of instances of message $m_i$ served is
\begin{align*}
	\sum_{\substack{k = 1}}^{j}
	\sum_{\substack{s = 1}}^{B}
	 \; r_k.B_s
	 \quad s.t. \; B_s = i
\end{align*}

\noindent
It may be that $m.o + m.d > m.p$, resulting in $\df(0)=-1$ (see \eqref{eq:df_def2}), like \eg in \figref{fig:afdfsf2}. This ``means'' that a message released at the each of one hyperperiod will have its deadline in the \emph{next} hyperperiod. 
To consider this situation, we introduce, for each message $m_i$, a variable $r_0.B_i$ set to the number of such ``leftover'' message instances at $t=0$. 
The system model makes the assumption that $\app.d \leq \app.p$. As $m.p = \app.p$, $m.o \leq m.p$ and $m.d < \app.d$, then $m.o + m.d < 2*m.p$. Thus, there can be only one or zero of such leftover message instances, \ie $r_0.B_i \in \{0\,,\,1\}$.
Finally, for each message $m_i \in \messageset$, and  $t \in \; [ \; r_{j} + \Tround \, ; \,  r_{j+1} + \Tround \; [$,
%
\begin{flalign}
\label{eq:sf_def2}
\sf_i: \; t \;
	&\longmapsto \;
	\sum_{\substack{k = 1 \\[2pt]s.t. \; r_k + \Tround \, < \, t}}^{j}\;\;
	\sum_{\substack{s = 1 \\[2pt]s.t. \; B_s = i}}^{B}
	 r_k.B_s - r_0.B_i 
\end{flalign}

\noindent
Ultimately, \textbf{(C4.1)} and \textbf{(C4.2)} can be formulated as ILP constraints using the following four equations:
\begin{flalign}
\notag
\forall\, m_i \in \messageset, \; \forall\, j\in [1..R],
&&&\eqref{eq:ka2}\quad \;	 : \;
	\quad	
	0 \, \leq \, r_j.t - m_i.o - (k^a_{ij}-1)*m_i.p \,<\, m_i.p 
&&\\
\notag
&&&\eqref{eq:kd2}\quad	\; : \;
	\quad	
	0 \, \leq \, r_j.t + \Tround - m_i.o - m_i.d - (k^d_{ij}-1)*m_i.p \,<\, m_i.p
&&\\
&&&\eqref{eq:af_const2}\quad	 \Leftrightarrow
	\quad	
	\sum_{k = 1}^j
	\sum_{\substack{s = 1 \\[2pt]s.t. \; B_s = i}}^{B}
	 \; r_k.B_s - r_0.B_i \; \leq\;  k^a_{ij}
&&\\
&&&\eqref{eq:df_const2}\quad	 \Leftrightarrow
	\quad	
	\sum_{k = 1}^{j-1}
	\sum_{\substack{s = 1 \\[2pt]s.t. \; B_s = i}}^{B} \; r_k.B_s - r_0.B_i 
	\; \geq\; 
	k^d_{ij}
&&
\end{flalign}

To implement those constraints in practice, some slight modifications are still required. We use Gurobi to solve the ILP problem. This solver does not allow to model strict inequalities, \ie $A \cdot x  < B$. Therefore, to implement the right-hand side of the constraints \eqref{eq:ka} and \eqref{eq:kd2}, we need to use a ``small'' time constant $mm$.
Furthermore, 
This leads to the following implementation. 

\begin{description}
	\item[(C4.1)]Every message must be served after its release time.
\end{description}
\begin{flalign}
&\forall\, m_i\in \messageset, \; \forall\, j\in [1..R_{\mode{}}],
\label{eq:serveAfterArrival}
&	&0  \; \leq \; r_j.t - m_i.o - (k^a_{ij}-1)*m_i.p 
		\; \leq \; m_i.p - mm
&		\\
&
&	&\sum_{k = 1}^{j}
	\sum_{\substack{s = 1 \\[2pt]s.t. \; B_s = i}}^{B} \; r_k.B_s - r_0.B_i 
	\; \geq\; 
	k^d_{ij}
&
\intertext{%
\begin{description}
	\item[(C4.2)]Every message must be served before its deadline.
\end{description}
}
&\forall\, m_i\in \messageset, \; \forall\, j\in [1..R_{\mode{}}],
&	&mm  \; \leq \; r_j.t + \Tround - m_i.o - m_i.d - (k^d_{ij}-1)*m_i.p 
		\; \leq \; m_i.p
&		\\
&
&	&\sum_{k = 1}^{j-1}
	\sum_{\substack{s = 1 \\[2pt]s.t. \; B_s = i}}^{B} \; r_k.B_s - r_0.B_i 
	\; \geq\; 
	k^d_{ij}
&
\end{flalign}

Finally, the formulation of the last two constraints -- \textbf{(C4.3)} and \textbf{(C4.4)} -- is rather straightforward.

\begin{description}
	\item[(C4.3)]A round cannot be allocated more messages than the number of slots available in one round.
\end{description}
\begin{align*}
\text{Satisfied by construction.}
\end{align*}

\begin{description}
	\item[(C4.4)] Within one hyperperiod, the same number of messages are released and served.
\end{description}			
\begin{flalign}
&\forall\, m_i\in \messageset,
&
	&\sum_{k = 1}^{R_{\mode{}}}
	\sum_{\substack{s = 1 \\[2pt]s.t. \; B_s = i}}^{B} \; r_k.B_s 
	\; = \; LCM/m_i.p
&&
\end{flalign}

\subsection*{Objective function}

To obtain a valid schedule, the ILP solver does not \emph{need} to optimize any objective function. The primer objective is to minimize of the number of rounds $R_{\mode{}}$ used in the schedule, which is achieved by incrementally increasing the number of rounds (see Algorithm~\ref{alg:outerlayer}) until a valid schedule is found.

However, from the application perspective, it is of importance to \emph{minimize the end-to-end latency} of concurrently running applications. Therefore, we formulate as an objective function, denoted \obj, the sum of all application latency, which we want to minimize. If we denote the latency of application \app by $\app.\delta$ and the latency of the chain $\app.c$ by $\app.c.\delta$,
\begin{align}
\intertext{%
$	\forall\, \app, \; 
	\forall\, \app.c \in \app.\predG, \;  
		\tau_{\first} = \app.c(\,\first\,), \; 
		\tau_{\last} = \app.c(\,\last\,), \;
		C = |\app.c|, $}
\app.c.\delta \; & = \; 
			\tau_{\last}.o + \tau_{\last}.e - \tau_{\first}.o + \sum_{j=1}^{C-1} \app.p*				\sigma_{i,j}
\\
 \app.\delta \; &= \; 
	\max_{\app.c \,\in\, \app.\predG}
		\left( 
			\;
			\app.c.\delta
			\;
		\right)
\intertext{%
And finally,}
\obj \; &= \; \sum_{\app \,\in\, \text{mode }\mode{}} \; 
	\app.\delta
\end{align}

\begin{table}[h]
\centering
\small
\caption{Complete list of variables used in the ILP formulation.}
\label{tab:ILPvar}
%
%
\begin{tabular}{		l			c		@{\qquad}	c	@{\qquad}			r		@{\,}		c		@{\,}			l		@{\,}	l	}
	\toprule																								
		Name	&		Notation		&	Type	&					&		Value		&					&		\\
		\toprule																							
		Task offset	&	$	\tau.o	$	&	Continuous	&	$	0	\leq	$	&	$	\tau.o	$	&	$	<	\tau.p	$	&		\\
		Message offset	&	$	m.o	$	&	Continuous	&	$	0	\leq	$	&	$	m_i.o	$	&	$	<	m.p	$	&		\\
		Message deadline	&	$	m.d	$	&	Continuous	&	$	0	\leq	$	&	$	m_i.d	$	&	$	\leq	m.p	$	&		\\
		\quad -	&	$	\sigma	$	&	Binary	&					&		0 \, or \, 1		&					&		\\
		\quad -	&	$	\lambda	$	&	Binary	&					&		0 \, or \, 1		&					&		\\
		Round starting time	&	$	r.t	$	&	Continuous	&	$	0	\leq	$	&	$	r_j.t	$	&	$	\leq	LCM - C_{net}	$	&		\\
		Round allocation	&	$	r.[B]	$	&	Integer	&	$	0	\leq	$	&	$	r_j.B_s	$	&	$	\leq	1	$	&		\\
		\quad -	&	$	r_0.B_i	$	&	Integer	&	$	0	\leq	$	&	$	r_0.B_i	$	&	$	\leq	1	$	&		\\
		\quad -	&	$	k^a_ {i,j}	$	&	Integer	&	$	0	\leq	$	&	$	k^a_ {i,j}	$	&	$	\leq	LCM/m_i.p	$	&		\\
		\quad -	&	$	k^d_ {i,j}	$	&	Integer	&	$	-1	\leq	$	&	$	k^d_ {i,j}	$	&	$	\leq	LCM/m_i.p	$	&		\\
		Number of rounds	&	$	R_{\mode{}}	$	&	Constant	&	$	0	\leq	$	&	$	R_{\mode{}}	$	&	$	\leq	R_{max}	$	&		\\
		Round time length	&	$	\Tround	$	&	Constant	&					&		1		&					&	time unit	\\
		Number of slots per round	&	$	B	$	&	Constant	&					&		5		&					&	slots	\\
		Minimal inter-round time	&	$	T_{max}	$	&	Constant	&					&		30		&					&	time unit	\\
		``Big'' time constant	&	$	M\!M	$	&	Constant	&					&		10*LCM		&					&		\\
		``Small'' time constant	&	$	mm	$	&	Constant	&					&		0.0001		&					&	time unit	\\
	\bottomrule																								
\end{tabular}

\end{table}

\end{document}